\input amstex
\documentstyle{amsppt}
\NoBlackBoxes
\nologo

\def \di{\partial}

\def \wt {\widetilde}

\def \mt {\mapsto}
\def \ra {\rightarrow}

\def \G {\Gamma}

\def \l {\lambda}
\def \L {\Lambda}

\def \S {\Sigma}

\def \t {\tau}
\def \o {\omega}

\def \LSl {\widetilde{\frak {Sl}}}

\def \CB {\bold{C}}

\def \IB {\bold{I}}

\def \DD {\Cal D}

\def \RR {\Cal R}

\def \di {\partial}

\def \ln{\text{ln}}
\def \tr{\text{tr}}

\def\ad {\text{ad}}
\def\Gr {\text{Gr}}

\def \diag{\text{diag }}

\def\maprightabove#1{\smash{\mathop{\longrightarrow}\limits^{#1}}}

\def\mapdownr#1{\Big\downarrow \rlap{$\vcenter{\hbox{$\scriptstyle#1$}}$}}
\def\mapdownl#1{\llap{$\vcenter{\hbox{$\scriptstyle#1$}}$}\Big\downarrow }

\hyphenation{non-au-to-nomous Sherbrooke Concordia Universit}
\topmatter
\title 
Bispectral Operators, Dual Isomonodromic Deformations \\
and the Riemann-Hilbert Dressing Method
\endtitle
\author 
J. Harnad
\endauthor
\leftheadtext{J. Harnad}
\rightheadtext{Bispectral Operators and Dual Isomonodromy}
\address
 Department of Mathematics and Statistics, Concordia University,
 7141 Sherbrooke W., Montr\'eal, Canada H4B 1R6, \ {\sl and}     
 Centre de recherches math\'ematiques, Universit\'e de Montr\'eal 
C.~P. 6128-A, Succ. centre--ville, Montr\'eal, Canada H3C 3J7
 \endaddress
\email harnad\@crm.umontreal.ca {\it or} harnad\@alcor.concordia.ca
\endemail

\vskip -0.5cm
\leftline{CRM-2512 (1997) 
\footnote{\eightpoint Text of talk presented at the workshop: {\it
The Bispectral Problem}, held at the Centre de recherches math\'ematiques,
Universit\'e de Montr\'eal, March 17--21, 1997. To appear in: CRM Proceedings 
and Lecture Notes series (1997/98).} 
\break}
\leftline{solv-int/9710016 \break} \bigskip
\vskip 0.5cm


\abstract
A comparison is made between bispectral systems and dual isomonodromic
deformation equations. A number of examples are given, showing how bispectral
systems may be embedded into isomonodromic ones. Sufficiency conditions are
given for the construction of rational solutions of isomonodromic deformation
equations through the Riemann-Hilbert problem dressing method, and these are
shown, in certain cases, to reduce to bispectral systems.
\endabstract

  \thanks Research supported in part by the Natural Sciences and
Engineering Research Council of Canada and the Fonds FCAR du Qu\'ebec.
\endthanks

\endtopmatter

\document

\head 1. Bispectral Systems vs. Isomonodromic Deformations
\endhead
We begin by listing a number of results characterizing bispectral systems
and isomonodromic systems, respectively, with a view to comparison.

\subhead 1.1. Bispectral Systems
\endsubhead

\smallskip
\roster
\item 
In bispectral systems, we have a function $\psi(x,z)$ of the two variables
$(x,z)$ that simultaneously satisfies a pair of eigenvalue equations
$$
\align
L\psi(x,z) &= f(z)\psi(x,z)  \tag{1.1a}\\
\L\psi(x,z) &= \phi(x)\psi(x,z),  \tag{1.1a}
\endalign
$$
where $L(x)$ and $\L(z)$ are differential operators in the indicated variables
and $f(z)$, $\phi(x)$ are parametric families of eigenvalues depending on the
other variable. Moreover \cite{DG}, for a suitable normalization,  $L(x)$ takes
the form
$$
L(x) = \di_x^r +a_{r-2}\di_x^{r-2} + \cdots  \tag{1.2}
$$
where the coefficients $\{a_j(x)\}_{j=1 \cdots r-2}$ are rational in $x$ 
and the eigenvalues $f(z)$ depend polynomially on $z$, and similarly for the
operator $\L(z)$ and eigenvalues $\phi(x)$.
 
\item
For $L$'s belonging to rank--$1$ bispectral algebras \cite{W}, the eigenfunction
$\psi$ may be viewed as the ${\bold t} = (t_1=x, 0, \dots)$ value of a
Baker-Akhiezer function $\psi_W(\bold{t},z)$ expressed, as usual, in terms of a
$\t$--function via the formula 
$$
\align
\psi_W& = e^{\sum_{m=1}^\infty t_m z^m} \t_W(\bold{t} -[\bold{z}]) \over
\t_W(\bold{t})
\tag{1.3a}\\
[\bold {z}] &:=\biggl({1\over z},  {1\over 2z}, {1\over 3z^3}, \cdots\biggr).
\tag{1.3b}
\endalign
$$
The $\tau$--function $\t_W$ is rational, and has the form
$$
\t_W = \det(X + \sum_{j=1}^\infty j t_j(-Z)^{j-1}),  \tag{1.4}
$$
where $(X,Z)$ is a pair of $n\times n$ matrices representing a point in Wilson's 
complexified, prereduced  Calogero--Moser phase space $C_n$. That
is, their commutator is of the form
$$
[X,\ Z] = -\IB + f g^T  \tag{1.5}
$$
for a pair of vectors $f, g\in \CB^n$. The $\bold{t}$--dependence of $\t_W$ is
thus given by flows generated by the commuting family of Hamiltonians
$$
h_n :=  -\tr(- Z)^n,  \tag{1.6}
$$
which are obtained by applying the bispectral involution (see below) to the
Calogero-Moser hierarchy. 
The Baker function $\psi_W$ at $\bold{t}=(x,0,\cdots)$ may be explicitly 
expressed in this case by the formula
$$
\psi_W(x,z) = e^{xz}\det(\IB_r - (X + x\IB)^{-1}(Z + z \IB)^{-1}). \tag{1.8}
$$

\item
Solutions can be constructed from suitably defined ``vacuum'' solutions
through the application of Darboux transformations \cite{BHY1, BHY2,K, K}.

\item 
The Baker function $\psi_W$ and the $\t$--function $\t_W$  for rank--$1$
bispectral algebras correspond to an element $W$ of Wilson's {\it adelic
Grassmannian} $\Gr^{\ad}$ \cite{W}. The existence of a bispectral pair $(L, \L)$
follows from the fact that $\Gr^{\ad}$ admits a {\it bispectral involution}
$b:\Gr^{\ad} \ra \Gr^{\ad} $ defined by the commuting diagram of invertible maps
$$
\matrix
\cup_n C_n & \maprightabove{b_W} & \cup_n C_n \\
\mapdownl{W} &  & \mapdownr{W} \\
\Gr^{\ad} & \maprightabove{b} & \Gr^{\ad}
\endmatrix  \tag{1.10}
$$
where the map $W:\cup_n  C_n \ra \Gr^{\ad}$ is defined by  formula
(1.8) and 
\hbox{$b_W:C_n \ra C_n$} by
$$
b_W : (X,Z) \mt (Z^t, X^t). \tag{1.11}
$$
\endroster

\subhead 1.2. Isomonodromic Deformations
\endsubhead 

\smallskip
\roster
\item
For these systems, we have an invertible $r \times r$ matrix--valued function
$\Psi(x,z)$ that simultaneously satisfies a pair of equations
$$
\align
{\di\Psi(x,z) \over \di x} &= U(x,z)\Psi(x,z) \tag{1.12a} \\
{\di\Psi(x,z) \over \di z} &= V(x,z)\Psi(x,z), \tag{1.12b}
\endalign
$$
where, in general, the $r \times r$ matrix--valued functions $U(x,z)$ and
$V(x,z)$ are rational in $z$. The consistency condition
$$
\left[{\di \over \di z} - U, \ {\di \over \di x} - V\right]  =0  \tag{1.13}
$$
implies the invariance of the (generalized) monodromy data of the
rational covariant derivative operator
$$
\DD_z:={\di \over \di z}-  V(x,z)  \tag{1.14}
$$
under the deformations resulting from varying $x$. In general, there may be
many deformation parameters $(t_1=x, t_2, \cdots )$, each with an associated
infinitesimal deformation operator
$$
\DD_j := {\di \over \di t_j} - U_j(t_1, t_2, \cdots, z) \qquad j=1,2, \dots  
\tag{1.15}
$$
commuting with $\DD_z$ and amongst themselves, each determining a
$1$--parameter family of isomonodromic deformations. The structure of the
corresponding deformation equations was analyzed in \cite{JMU, JM}.

\item
Isomonodromic systems of type (1.12a,b) have a Hamiltonian structure \cite{JMU,
H1, HTW} and are generated by certain spectral invariants of the matrix $V$. The
logarithmic differential of the corresponding $\t$--function is given by
$$
d(\ln \t) = \sum_{j} H_j dt_j,  \tag{1.16}
$$
where $\{H_j\}_{j=1,2, \cdots}$ are the spectral invariant Hamiltonians
generating the equations (1.12a,b).

\item
Solutions to eqs.~(1.12a,b) can be constructed from ``vacuum'' solutions
by application of the  ``dressing method'' \cite{NZMP, HI}, based on the
matrix Riemann-Hilbert problem. For suitably chosen vaccua and loop
group elements, this gives rise to solutions having a rational dependence upon
the deformation parameter.

\item
The isomonodromic $\t$--functions correspond to certain special elements $W\in
\Gr$ of the general Segal--Wilson--Sato Grassmannian.
Moreover, at least for the case where the element $V(x,z)$ has
$n$ simple poles at finite $z$ and tends to a nonsingular finite value $X$ at
$z=\infty$, there exists an equivalent representation of the underlying
dynamical equations as deformations of a second {\it dual} isomonodromic family 
of differential operators \cite{H1}
$$
\DD_{\l} := {\di \over \di \l} - \wt{V}(x,\l), \tag{1.17}
$$
with $\wt{V}(x,\l)$ tending to a nonsingular finite limit $Z^T$ at $\l=\infty$.
The corresponding {\it dual} matrix Baker function $\wt{\Psi}(x,\l)$
is a nonsingular $n \times n$ matrix satisfying a similar system of equations
$$
\align
{\di\wt{\Psi}(x,\l) \over \di x} &= \wt{U}(x,\l)\wt{\Psi}(x,\l) \tag{1.18a} \\
{\di\wt{\Psi}(x,\l) \over \di \l} &= \wt{V}(x,\l)\wt{\Psi}(x,\l), \tag{1.18b}
\endalign
$$
where the $n \times n$ matrices $\wt{U}(x,\l)$ and $\wt{V}(x,\l)$ are rational in
the second spectral parameter $\l$. The dual system (1.18a,b) can be related 
to the original one (1.12a,b) by introducing an auxiliary symplectic vector space
consisting of canonically conjugate pairs $(F.G)$ of $n\times r$ rectangular
matrices and applying a Hamiltonian quotienting procedure. The resulting
matrices $V$ and $\wt{V}$ in the definition of the operators $\DD_z$ and $\DD_\l$
respectively are given by the formulae
$$
\align
V(x,z) &=X + F(Z - z\IB_r)^{-1} G^T  \tag{1.19a}\\
\wt{V}(x,z) &=Z^T + F^T(X^T - \l\IB_n)^{-1}G, \tag{1.19b}
\endalign
$$
and the corresponding spectral curves are birationally equivalent (with the
r\^oles of the loop parameter $\l$ and eigenvalue $z$ interchanged). These are
therefore related by the involutive {\it duality} map

$$
 (X,Z,F,G,z, \l) \mt (Z^T, X^T, F^T, G^T, \l, z).  \tag{1.20}
$$
\endroster

\subhead 1.3. A simple example
\endsubhead 
  To further indicate the close correspondence between bispectral and
isomonodromic systems, consider the following example of rank--$1$ bispectral
systems (cf.~\cite{H2}). In the above notation, choose the matrices $X$ and $Z$
to be
$$
X= \pmatrix 0 & 1 \\ 0 & 0 \endpmatrix, 
\qquad Z= \pmatrix 0 & 0 \\ -1 & 0 \endpmatrix.  \tag{1.21}
$$
The resulting bispectral Baker function is then
$$
\psi(x,a) = e^{x z} \left(1 - {2\over xz} + {2\over x^2 z^2}\right) , \tag{1.22}
$$
which satisfies the eigenvalue equations
$$
\align
\left({\di^3 \over \di x^3} - {6\over x^2}{\di \over \di x} + {12 \over
x^3} \right) \psi &= z^3 \psi \tag{1.23a}  \\
\left({\di^3 \over \di z^3} - {6\over z^2}{\di \over \di z} + {12 \over
z^3}\right) \psi &= x^3 \psi \tag{1.23b}
\endalign
$$
In this case, we have two other single--valued solutions of the same system, due
to the invariance under multiplication of either $x$ or $z$ by a cube root of
$1$. Let  
$$
\align
\Psi &:=\pmatrix \psi_0 & \psi_1 & \psi_2 \\
                \psi_{0,x} & \psi_{1,x} & \psi_{2,x}  \\
                \psi_{0,xx} & \psi_{1,xx} & \psi_{2,xx}
        \endpmatrix
\tag{1.24a}  \\
\psi_j(x,z) &:= \psi(x, \o^j z),  \qquad \o := e^{2\pi i\over 3},
 \qquad j=0,1,2,
\tag{1.24b}
\endalign
$$
be the Wronskian matrix formed from the three solutions
so obtained.
Then $\Psi$ satisfies
$$
\align
{\di \Psi\over \di x} - 
\pmatrix 0 & 1 & 0  \\
      0 & 0 & 1  \\
   z^3 - {12\over x^3} & {6\over xz} & 0 
\endpmatrix \Psi =0   \tag{1.25a}\\
{\di \Psi\over \di z} - 
\pmatrix 0 & x \over z & 0  \\
      0 & 1\over z & x \over z  \\
   x z^2 - {12\over x^2 z} & {6\over x z} & 2\over z 
\endpmatrix \Psi=0, \tag{1.25b}
\endalign
$$
which shows that, viewing $x$ as a deformation parameter, the monodromy of the
rational covariant derivative operator entering in (1.25b) is independent of
$x$. Of course, the monodromy in this case is actually trivial, since the
matrix Baker function $\Psi$ is globally defined and single--valued. Moreover
since the dependence on {\it both} the variables $x$ and
$z$ is rational, the system could equally well be viewed as determining a
$1$-parameter family of rational covariant differential operators in the $x$
variable, defined by eq.~(1.25a), having trivial monodromy, with $z$ viewed
as the deformation parameter. A {\it dual} isomonodromic system
may also be defined, just by introducing the dual matrix Baker function
$\wt{\Psi}$ as the Wronskian matrix determined  from $(\wt{\psi}_0, \wt{\psi}_1,
\wt{\psi}_2)$ through differentioation with respect to the $z$--variable,
$$
\align
\wt{\Psi} &:=\pmatrix \wt{\psi}_0 & \wt{\psi}_1 & \wt{\psi}_2 \\
                \wt{\psi}_{0,z} & \wt{\psi}_{1,z} & \wt{\psi}_{2,z}  \\
                \wt{\psi}_{0,z} & \wt{\psi}_{1,z} & \wt{\psi}_{2,z}
        \endpmatrix
\tag{1.26a}  \\
\wt{\psi}_j(x,z) &:= \psi(x, \o^j z),  \qquad \o := e^{2\pi i\over 3}, \qquad j=0,1,2
\tag{1.26b}
\endalign
$$
This then satisfies the system obtained from (1.25a,b) by interchanging the
$x$ and $z$ variables.

  The question naturally arises: {\it is it possible in general to embed
bispectral systems into isomonodromic ones, such that the matrix
Riemann-Hilbert problem used in the dressing method construction of solutions to
the latter systems reduces to the Darboux transformations
used in constructing the former}?  In the following sections, this
idea will be further developed. A set of criteria is given, within
the Riemann-Hilbert problem setting, giving rise to isomonodromic systems
derived from suitably defined ``vacuum'' solutions. Imposing further
restrictions on the data entering the Riemann--Hilbert problem, we deduce
sufficient conditions for the solution  to depend rationally on the deformation
parameter. Although a set of sufficient conditions for these systems to reduce
to  bispectral ones is not yet determined, we are able to illustrate via
examples and elementary calculations that rank--$1$ bispectral systems can be
recovered from this approach

\head 2. Isomonodromic Systems and the Riemann--Hilbert Problem
\endhead

\subhead 2.1. The Dressing Method for Isomonodromic Systems
\endsubhead
   In the matrix Riemann-Hilbert (RH) problem approach to 
zero--curvature equations depending upon a spectral parameter $z$,
one usually starts with a particular $r \times r$ matrix function 
$\Psi_0(\bold{t},z)$, viewed as the ``vacuum'' solution, depending on a number 
of commuting flow parameters $\bold{t} = (t_1, t_2, \cdots)$ and chooses, along
some suitably defined, closed contour $\G$ in the $z$--plane, with interior
region
$\G_+$ and exterior $\G_-$,  a smooth, nonsingular $r \times r$ matrix--valued
function $H(z)$ of the spectral parameter. (In the group theoretical
formulation, $H$ is interpreted as a loop group element.) The Zakharov--Shabat 
{\it dressing method} \cite{NZMP} then consists of solving the associated matrix
RH problem for the conjugated matrix
$$
\chi_+^{-1}(\bold{t},z) \chi_-(\bold{t},z)=\Psi_0(\bold{t},z) H(z)
\Psi_0^{-1}(\bold{t},z), 
\tag{2.1}
$$
where $\chi_{\pm}$ are holomorphic functions of $z$ in the regions $\G_{\pm}$
which, in the regular case, are nonsingular $r \times r$ matrices in their domain
of holomorphicity (with a suitable normalization condition at some point, usually
$z=\infty$, to guarantee uniqueness of the solution). The ``dressed'' matrix
Baker function $\Psi(x,z)$ is then obtained by applying $\chi_{\pm}$ as a gauge
transformation to the vacuum solution
$$
\Psi = \chi_+ \Psi_0 H_+ = \chi_- \Psi_0 H_-^{-1} , \tag{2.2}
$$
where the nonsingular $r \times r$ matrices $H_{\pm}(z)$ are
$\bold{t}$--independent and chosen in any way that respects the factorization
$$
H(z) = H_+(z) H_-(z).  \tag{2.3}
$$
These just serve to determine a basis for the fundamental
solution and may, in particular, be chosen either so that eq.~(2.3) itself is a
solution to the RH problem for the fixed element $H$ or, alternatively, that
one of the two factors $H_{\pm}$ is equal to $H$ itself, and the other is
the identity matrix.

If the vacuum  matrix Baker function satisfies the Zakharov--Shabat (ZS)
system
$$
{\di \Psi_0 \over \di t_j} = U_{j,0}(\bold{t},z) \Psi_0,   \tag{2.4}
$$
where the $U_0^j$'s are $r \times r$ matrix--valued functions depending
rationally on the spectral parameter $z$, with a given pole structure and
asymptotic limit as $z\ra\infty$, the dressed Baker function
$\Psi(\bold{t},z)$ will satisfy a system having the same pole and asymptotic
structure in $z$
$$
{\di \Psi \over \di t_j} = U_j(\bold{t},z) \Psi_0,   \tag{2.5}
$$
where the $U_j$'s are obtained by applying the same gauge transformation to the
vacuum covariant derivative operators as that appearing in eq.~(2.2). It
follows that the compatibility conditions for (2.4), the
zero--curvature conditions, are satisfied identically in $z$.

  To adapt this procedure to isomonodromic deformation equations, we must
assume that, in addition to the usual vacuum ZS equations (2.4), $\Psi_0$ also
satisfies  a compatible differential equation with respect to the spectral
parameter
$$
{\di \Psi_0\over \di z} = V_0 (\bold{t}, z) \Psi_0,  \tag{2.6}
$$
where the $r\times r$ matrix--valued function $V_0$ depends polynomially on $z$.
This implies that the (generalized) monodromy data of the vacuum rational
covariant derivative operator
$$
\DD^0_z := {\di \over \di z} - V_0(\bold{t},z) \tag{2.7}
$$
is invariant under the deformations generated by varying the parameters 
$(t_1,t_2, \cdots )$. In order to guarantee that the resulting ``dressed''
operator
$$
\DD_z := {\di \over \di z} - V(\bold{t},z), \tag{2.8}
$$
obtained through application of the gauge transformation (2.2) to $\DD^0_z$,
should also be rational in the spectral parameter $z$, further restrictions must
be put on the choice of $H(z)$.  A sufficient set of conditions for $H(z)$ is
given in the following.
\proclaim{Proposition 2.1} If $H(z)$ satisfies a differential equation of the
form
$$
{d H(z) \over d z} = r(z) H(z)  + Q(z),  \tag{2.9}
$$
where $r(z)$ is a rational matrix--valued function and $Q(z)$ is a distribution
with support at a finite number of points, and the splitting (2.3) is chosen
such that
$$
H_+(z) := H(z), \qquad H_-(z)= \IB,  \tag{2.10}
$$
the resulting operator $\DD_z$ will depend rationally on $z$, and its
(generalized) monodromy will be preserved under deformations in the parameters 
$\{t_j\}_{j=1,2 \cdots}$ determined compatibly by the equations (2.5).
If $Q(z)$ vanishes, then $V(\bold{t},z)$ can be expressed
$$
V(\bold{t},z) = V_-(\bold{t},z) + V_+(\bold{t},z), \tag{2.11}
$$
where $V_+$ is a polynomial in $z$, of the same degree as $V_0$, and $V_-$ is
rational in $z$, with the same pole support as $r(z)$, and vanishing at
$z=\infty$.
\endproclaim

The proof of this proposition is quite elementary and will not be detailed
here. However, the sense of the derivative appearing in eq.~(2.8) needs to be
clarified since, {\it a priori}, $H(z)$ is only defined along the contour
$\G$. What is meant here is that we assume that $H(z)$ may be extended to a
neighborhood of $\G$, though not necessarily as a holomorphic function. If it
can be extended holomorphically, terms like $Q(z)$ are absent. In
particular, $H(z)$ could be piecewise constant along $\G$, and $Q(z)$ could
have its support just on $\G$ itself. This case will not be the one of
interest to us in the following, but it does play an important r\^ole in the
solution of certain classes of isomonodromic deformation equations through the
Riemann-Hilbert problem method (cf. \cite{HI}).

  In the following, we shall restrict ourselves to the case when
$Q(z)\equiv 0$, and seek to impose suitable restrictions on the vacuum solution
$\Psi_0$ and the element $H$ in order that the resulting matrices
$(U_j(\bold{t}, z), V(\bold{t}, z))$ appearing in eqs.~(2.5) and (2.8) 
depend rationally, not only on the spectral parameter $z$, but also on the
deformation parameters $\{t_j\}$. We begin in the next subsection by giving two
examples that lead to such rational solutions, and which may be reduced to
particular cases of Wilson's rank-$1$ bispectral systems.

\subhead 2.2. Two Examples Reducing to Rank-$1$ Bispectral Systems 
\endsubhead

The following examples show how an appropriately chosen pair $(\Psi_0, \
H)$  leads, through solution of the Riemann-Hilbert problem, to isomonodromic
systems that reduce to bispectral ones.  We restrict ourselves to
$3 \times 3 $  matrices, with $\chi_-$ rational, and the pole support of both
$\chi_-$ and $H_z H^{-1}$ at $z=0$. To obtain a correspondence between the
scalar and matrix systems, we restrict ourselves to the subgroup of the full loop
group $\LSl(3)$ consisting of elements satisfying the invariance condition
$$
T H(\o z) T^{-1} =  H(z)  \tag{2.12}
$$
where
$$
\o := e^{2\pi i\over 3}, \qquad
T:= 
\pmatrix
1 & 0 & 0 \cr
0 & \o & 0 \cr
0 & 0 & \o^2
\endpmatrix, \tag{2.13}
$$
with similar rectrictions on $\Psi_0$, $V_0$ and $U_0$. Define the matrix
$$
\S:= 
\pmatrix
0 & 0 & 1 \cr
1 & 0 & 0 \cr
0 & 1 & 0
\endpmatrix. \tag{2.14}
$$
In both the following examples we choose the vacuum Baker function as
$$
\Psi_0(x,z) := e^{x z \S},  \tag{2.15}
$$
so that
$$
U_0:= z \S \qquad
V_0:= x \S.   \tag{2.16}
$$
From any $3\times 3$ matrix Baker function $\Psi$ satisfying the
invariance condition (2.12), we obtain three scalar Baker functions
 by summing along the rows; i.e., by taking the components of the vector
obtained by applying $\Psi$ to the vector with all entries equal to $1$.
$$
\pmatrix \psi_1 \cr \psi_2 \cr \psi_3 \endpmatrix
:= \Psi\pmatrix 1 \cr 1 \cr 1 \endpmatrix. \tag{2.16}
$$

\example{Example 2.2.1}
 Choose 
$$
H(z) := \pmatrix
0 & z & 0 \cr
0 & 0 & z \cr
z^{-2} & 0 & 0
\endpmatrix  \tag{2.17}
$$
as the loop group element.
The solution to the Riemann-Hilbert problem is then given by
$$
\chi_- = \pmatrix
1 & 0 & 0 \cr
-{1\over xz} & 1 & 0 \cr
{2\over x^2 z^2} &-{2\over x z} & 1
\endpmatrix , \tag{2.18}
$$
which is rational in $z$.
The corresponding isomonodromic system (1.12a,b), (1.13) is then
defined by
$$
U(x,z) =
\pmatrix
{1\over x} & z  & 0 \cr
0 & {1\over x} & z \cr
z  & 0 & -{2\over x}
\endpmatrix ,  \qquad
V(x,z) =
\pmatrix
{1\over z} & x  & 0 \cr
0 & {1\over z} & x \cr
x  & 0 & -{2\over z}
\endpmatrix .
 \tag{2.19}
$$
Summing over the rows of $\Psi$ gives
$$
\psi_1 = e^{xz}, \quad
\psi_2 = e^{x z} \biggl( 1 - {1\over x z}\biggr), \quad
\psi_3 = e^{x z} \biggl( 1 - {2\over x z} + {2\over x^2 z^2}\biggr).
\tag{2.20}
$$
These are all rank-$1$ bispectral wave functions which are eigenfunctions,
respectively, of the following pairs of operators
$$
\align
L_1 &= {\di \over \di x}, \qquad \L_1 = {\di \over \di z},  \tag{2.21a} \\
L_2 &= {\di^2 \over \di x^2} -{2\over x^2}, \qquad 
\L_2 ={\di^2 \over \di z^2} -{2\over z^2},  \tag{2.21b} \\
L_3 &= {\di^3 \over \di x^3}  -{6\over x^2} {\di \over \di x} + {12\over
x^3},
\qquad 
\L_3 = {\di^3 \over \di z^3}  -{6\over z^2} {\di \over \di z} + 
{12\over z^3}.
\tag{2.21c}
\endalign
$$
In terms of Wilson's correspondence, $\psi_1$ is just the vacuum Baker
function, $\psi_2$ corresponds to the  point in the $1$--particle Calogero-Moser
phase space with $(X,Z)=(0,0)$, and $\psi_3$ is just the example of
section 1.3, corresponding to the point in the $2$--particle phase space with
matrices $(X, \ Z)$ given in (1.21). Note, however, that the correspondence
between the isomonodromic and bispectral systems given here is not the
same as the one given in section (1.3).
\endexample

\example{Example 2.2.2}
Now choose
$$
H(z) := \pmatrix
0 & 0 & z^2 \cr
z^{-1} & 0 & 0 \cr
0 & z^{-1} & 0
\endpmatrix.  \tag{2.23}
$$
The solution to the Riemann-Hilbert problem in this case gives
$$
\chi_- = \pmatrix
1 & 0 & 0 \cr
-{2\over xz} & 1 & 0 \cr
0 &-{1\over x z} & 1
\endpmatrix , \tag{2.24}
$$
and the corresponding isomonodromic system (1.12a,b), (1.13) is 
defined by
$$
U(x,z) =
\pmatrix
{2\over x} & z  & 0 \cr
0 & -{1\over x} & z \cr
z  & 0 & -{1\over x}
\endpmatrix ,  \qquad
V(x,z) =
\pmatrix
{2\over z} & x  & 0 \cr
0 & -{1\over z} & x \cr
x  & 0 & -{1\over z}
\endpmatrix .
 \tag{2.25}
$$
The associated bispectral Baker functions obtained by summing over the
rows of $\Psi$ are
$$
\psi_1 = e^{xz}, \quad
\psi_2 = e^{x z} \biggl( 1 - {2\over x z}\biggr), \quad
\psi_3 = e^{x z} \biggl( 1 - {1\over x z}\biggr),
\tag{2.26}
$$
so $\psi_1$  and$\psi_3$ coincide with as $\psi_1$  and $\psi_3$, respectively,
of the previous example, while $\psi_2$ here is a bispectral eigenfunction of the
pair of operators
$$
L = {\di^3 \over \di x^3}  -{6\over x^2} {\di \over \di x},  \qquad
\L= {\di^3 \over \di z^3}  -{6\over z^2} {\di \over \di z}.  \tag{2.27} 
$$
\endexample
This case corresponds to the point in the Calogero-Moser phase space with
matrices
$$
X = \pmatrix 0 & 1 \cr 0 & 0 \endpmatrix, \qquad
Z = \pmatrix 0 & 0 \cr 1 & 0 \endpmatrix.  \tag{2.28}
$$

  In both these examples, $\chi_-$ is rational, and its only pole,  at $z=0$,
is coincident with that of $H_z H^{-1}$. In the last subsection,  some
preliminary results will be given showing how isomonodromic systems may more
generally be obtained, such that the dependence on the deformation parameter is
also rational, but with an arbitrary number of poles in $r(z)$. The
structure of the solutions wil be seen to closely resemble that of bispectral
systems.
\subhead 2.3. Rational Isomonodromic Systems
\endsubhead

We now make some more restrictive assumptions about the solution to the
Riemann-Hilbert problem, which are sufficient to assure that the matrices
$U_j(\bold{t},z)$ and $V(\bold{t},z)$ are rational functions of
the deformation parameters $(x=t_1, t_2 \cdots )$. In the following, we
assume that the contour $\G$ is sufficiently large that all poles of $r(z)$ are
contained in the interior region $\G_+$, and choose the normalization of
$\chi_-(z)$  to be such that
$$
\chi_-(\infty) =\IB.  \tag{2.29}
$$
We must of course first require that the matrices
$U_{0,j}(\bold{t}, z)$, $V_0(\bold{t}, z)$ entering in the {\it vacuum} 
equations (2.5), (2.6) should themselves depend rationally on the deformation
parameters.  Now also assume that the solution to the Riemann-Hilbert problem
is such $\chi_-$ is {\it rational} in $z$. It follows that its pole support is
contained in that of $r(z)$. We now add the additional requirement that the
difference
$$
\chi_+(z) \Psi_0(z) r(z) \Psi_0^{-1}(z) \chi_+^{-1}(z) -r(z) \tag{2.30}
$$
be holomorphic in $\G_+$. This means, for example, when $r(z)$ has only first
order poles at the points $z= z_1, \cdots z_n$, with residue matrices
$\{r_i\}_{i=1\cdots n}$, that the matrix $\chi_+(z_i)\Psi_0(z_i)$ must
commute with $r_i$, for each $z=z_i$. We then have the following result

\proclaim{Proposition 2.2} Under the above assumptions, (i.e., that $\chi_-$,
$U_{0,j}$ and $V_0$ are all rational in the deformation parameters, and
$\chi_+(z)\Psi_0(z)$ is such that the expression (2.30) is holomorphic in $z$
inside $\G_+$), the matrices $U_j(\bold{t},z)$, $V(\bold{t},z)$ obtained
through the dressing transformation are also rational in the parameters
$\bold{t}=(t_1, t_2 \cdots )$.
\endproclaim
Again, the proof is elementary and will not be detailed here. However, to show
the close similarity with bispectral systems, at least for the case of rank--$1$
bispectral algebras, we will consider the detailed  structure of the
resulting Baker matrix $\Psi$ in the case when $r(z)$ has only simple poles and
$V_0(x,z)$ is chosen as independent of $z$. (Here $x=t_1$ and the other
deformation parameters are chosen to vanish.) This means that the vacuum Baker
matrix $\Psi_0$, as in the scalar case, is a linear exponential in $z$, with
matrix exponent $z V_0(x)$. (These latter restrictions are imposed only for the
sake of simplicity and comparison with the bispectral case; the following
formulae can easily be modified to cover the general case with higher
order poles and polynomial dependence on $z$.) For this case, $r(z)$ and
$\chi_-(x,z)$ have the general form
$$
\align
r(z) &= \sum_{j=1}^n {r_i\over z -z_i},  \tag{2.31a} \\
\chi_-(x,z) &= \IB + \sum_{j=1}^n {Q_i \over z- z_i},  \tag{2.31b}
\endalign
$$
where the residue matrices $\{r_j\}$ are constants and the $Q_i$'s are
rational functions of $x$, whose explicit form will be determined under the
assumptions of Proposition 2.2. The relation (2.29) in this case implies the
equalities
$$
\chi_+(z_i)\Psi_0(z_i) r_i \Psi_0^{-1}(z_i) \chi_+^{-1}(z_i) =r_i,
\quad i =1 , \cdots n.  \tag{2.32}
$$
It then follows from the dressing method that the polynomial (in $z$) and
rational parts $V_+$, $V_-$, respectively, of the matrix $V(x,z)$ entering in
Proposition 2.1 are just
$$
V_+(x,z) = V_0(x), \qquad V_-(x,z) = r(z).  \tag{2.33}
$$
Thus, under the given assumptions, the residue matrices in the polar part of
$V(x,z)$ are constants, and the polynomial part is the same is for the vacuum
solution. The dressing transformation gives
$$
\bigl(V_0(x) + r(z)\bigr) \chi_- (x,z) = \chi_{-,z}(x,z) + \chi_-(x,z) V_0(x).
\tag{2.34}
$$
Substituting the expression (2.31a,b) for $\chi_-(x,z)$ and $r(z)$ and
equating the polar parts gives the realations
$$
\align
\biggl( \ad_{V_0} - \sum_{j=1 \atop j\neq i}^n {r_j \over z_i -z_j}\biggr) Q_i
+ \sum_{j=1 \atop j\neq i}^n {r_i \over z_i -z_j} Q_j& = - r_j \tag{2.35b}  
\cr
r_j Q_j& = - Q_j,   \quad j=1 \cdots n.  \tag{2.35b}
\endalign
$$
To express the solution of this linear algebraic system it is convenient to
introduce the $nr \times r$ matrices formed from a column of $r\times r$
blocks consisting of the residue matrices
$$
R:= \pmatrix r_1 \cr . \cr . \cr r_n \endpmatrix,  \qquad
Q:= \pmatrix Q_1 \cr . \cr . \cr Q_n \endpmatrix,   \tag{2.36}
$$
as well as the $nr \times nr$ matrix $X$, consisting of $r\times r$
blocks $\{X_{ij}\}_{i,j =1,\cdots n}$ given by
$$
\align
X_{ii} &:= - \sum_{j=1 \atop j\neq i}^n {r_j \over z_i -z_j}   \tag{2.37a}
 \cr
X_{ij} &:=  {r_j \over z_i -z_j}, \quad i\neq j. \tag{2.37b}
\endalign
$$
Eq.~(2.35a) may then be written more concisely as
$$
\biggl( \ad_{V_0} \otimes \IB + X\biggr) Q = -R.  \tag{2.38}
$$
Solving, and substituting into the expression (2.31b) for $\chi_-$  and (2.2)
for $\Psi$ gives
$$
\Psi(x,z) = \biggl(\IB  + E^T(Z- z\IB )^{-1}(X + \ad_{V_0} \otimes \IB)^{-1}R
 \biggr), \tag{2.39}
$$
where $E$ is the $nr \times r$ matrix consisting of a column of $n$ 
blocks, each of which is the $r\times r$ identity matrix
$$
E = \pmatrix \IB \cr . \cr . \cr \IB \endpmatrix,  \tag{2.40}
$$
while $Z$ is the $np \times np$ block--diagonal matrix whose $r\times r$
diagonal blocks are just $- z_1 \IB$
$$
Z = - \diag(z_1, z_2, \cdots z_n) \otimes \IB.  \tag{2.41}
$$
The commutator of the matrices  $X$ and $Z$ is given by
$$
[X, \ Z] = -\RR + E R^T,  \tag{2.42}
$$
where $\RR$ is the $nr \times nr$ block diagonal matrix with $j$'th $r \times r$
diagonal block equal to $R_j$. 

   We see that the rational $x$--dependence
contained in formula (2.39) comes entirely from the factor 
$(X + \ad_{V_0} \otimes \IB)^{-1}$. Comparison with the formulae in \cite{W}
for the Baker functions for rank--$1$ bispectral algebras shows the close
resemblance to eq.~(2.39), while eq.~(2.42) is the analogue of the relation
(1.5) satisfied by the associated Calogero--Moser matrices.

  Although the above scheme does not guarantee reduction to scalar bispectral
systems, the examples of the previous subsection show that they  do so at
least in particular cases. Also, although the conditions of Proposition (2,2)
can be used constructively when looking for isomonodromic systems involving
rational dependence on the deformation parameters, it would be more satisfying
to find sufficient conditions for rationality, and for bispectrality,
purely in terms of the input data consisting of the vacuum solution $\Psi_0$
and the group element $H$.

\bigskip\bigskip
\noindent{\it Acknowledgements.} The author would like to thank
 A. Kasman and A. Orlov for helpful discussions. 
\bigskip\bigskip  
\refstyle{A}


\widestnumber\key{BHY1}
\Refs

\ref\key{BHY1}
\by  B. Bakalov, E. Horozov, M. Yakimov
\paper B\"acklund--Darboux transformations in Sato's Grassmannian
\jour preprint (Sofia Univ.), q-alg/9602010
\vol 
\yr 1996
\pages   
\endref

\ref\key{BHY2}
\by  B. Bakalov, E. Horozov, M. Yakimov
\paper Highest weight modules of $W_{1+\infty}$, Darboux transformations and the
bispectral problem
\jour preprint (Sofia Univ.), q-alg/9601017
\vol 
\yr 1996
\pages   
\endref

\ref\key{DG}
\by  J.J. Duistermaat and A. Gr\"unbaum
\paper Differential equations in the
Spectral parameter
\jour Commun. Math.  Phys.
\vol 107
\yr 1986
\pages 177--2404
\endref

\ref\key{H1}
\by  J. Harnad
\paper Dual Isomonodromy Deformations and Moment Maps 
to Loop Algebras
\jour Commun. Math.  Phys.
\vol 166
\yr 1994
\pages 337--365
\endref

\ref\key{H2}
\by  J. Harnad
\paper Bispectral Operators of Rank--$1$ and Dual Isomonodromic Deformations 
\jour Preprint CRM 2443 (1996), solv-int/9612012, to appear in 
AMS-CRM Conference and Lecture Note Series {\bf 21}  
\vol 
\yr 1996
\pages 
\endref

\ref\key{HI}
\by  J. Harnad and A. Its
\paper Integrable Fredholm Operators and
       Dual Isomonodromic Deformations  
\jour Preprint CRM 244x (1997), solv-int/9712012  
\vol 
\yr 1997
\pages 
\endref

\ref\key{HTW}
\by J. Harnad, C. Tracy, and H. Widom
\paper Hamiltonian Structure of
Equations  Appearing in Random Matrices
\inbook NATO ASI Series  B, vol. {\bf 314},
 {\it Low Dimensional Topology and Quantum Field Theory},
ed. H. Osborn
\publ Plenum 
\publaddr New York
\yr 1994
\pages 231--245
\endref

\ref\key{JMU}
\by  M. Jimbo, T. Miwa, and K. Ueno
\paper`Monodromy Preserving Deformation of Linear Ordinary Differential 
Equations with Rational Coefficients I.
\jour Physica 
\vol 2D
\yr 1981
\pages  306--352
\endref

\ref\key{JM}
\by  M. Jimbo and T. Miwa
\paper Monodromy Preserving
Deformation of Linear Ordinary Differential Equations with Rational
Coeefficients II, III
\jour Physica 
\vol 2D, 4D
\yr 1981
\pages   407-448,  26--46 
\endref

\ref\key{K}
\by  A. Kasman
\paper Bispectral Darboux Transformations: The Generalized Airy Case
\jour  preprint (Univ. of Georgia mathematics), q-alg/9606018  (to appear in
Physica {\bf D})
\vol 4, no. 2
\yr 1996
\pages   
\endref

\ref\key{NZMP}
\by  S.~P. Novikov, V.~E. Zakharov, S.~V. Manakov,  L.~V. Pitaevski,
\book Soliton Theory: The Inverse Scattering Method
\publ Plenum
\publaddr New York
\yr 1984
\endref

\ref\key{W}
\by G. Wilson
\paper Bispectral Commutative Ordinary Differential Operators
\jour J. Reine Angew. Math.
\vol 442
\yr 1993
\pages 179--204
\endref

\endRefs
\enddocument